\documentclass[10pt]{manuscript}
\usepackage{cite}

\usepackage{soul}
\usepackage{amsmath}
\usepackage{subfig}
\usepackage{bm}
\usepackage{colortbl}
\usepackage{hyperref}
\usepackage{amsmath}
\usepackage{amssymb}
\usepackage{graphicx}
\usepackage{dsfont}
\usepackage{algorithm}
\usepackage{float}
\usepackage[noend]{algpseudocode}
\hypersetup{
    colorlinks=true,
    linkcolor=black,
    citecolor=black,
    filecolor=magenta,      
    urlcolor=cyan,
    pdftitle={Overleaf Example},
    pdfpagemode=FullScreen,
    }
\newcommand{\beginsupplement}{%
        \setcounter{table}{0}
        \renewcommand{\thetable}{S\arabic{table}}%
        \setcounter{figure}{0}
        \renewcommand{\thefigure}{S\arabic{figure}}%
     }

\title{COSMOS: A Data-Driven Probabilistic Time Series simulator for Chemical Plumes across Spatial Scales}

\author[a]{Arunava Nag}
\author[b,c,d]{Floris van Breugel}

\affil[a]{Computer Science Engineering Department, University of Nevada, Reno}
\affil[b]{Integrative Neuroscience Program, University of Nevada, Reno}
\affil[c]{Ecology Evolution and Conservation Biology Program, University of Nevada, Reno}
\affil[d]{Department of Mechanical Engineering, University of Nevada, Reno}

\corrauthor[1]{Floris van Breugel}{fvanbreugel@unr.edu}

\begin{abstract}
The development of robust odor navigation strategies for automated environmental monitoring applications requires realistic simulations of odor time series for agents moving across large spatial scales. Traditional approaches that rely on computational fluid dynamics (CFD) methods can capture the spatiotemporal dynamics of odor plumes, but are impractical for large-scale simulations due to their computational expense. On the other hand, puff-based simulations, although computationally tractable for large scales and capable of capturing the stochastic nature of plumes, fail to reproduce naturalistic odor statistics. Here, we present COSMOS (Configurable Odor Simulation Model over Scalable Spaces), a data-driven probabilistic framework that synthesizes realistic odor time series from spatial and temporal features of real datasets. COSMOS generates similar distributions of key statistical features such as whiff frequency, duration, and concentration as observed in real data, while dramatically reducing computational overhead. By reproducing critical statistical properties across a variety of flow regimes and scales, COSMOS enables the development and evaluation of agent-based navigation strategies with naturalistic odor experiences. To demonstrate its utility, we compare odor-tracking agents exposed to CFD-generated plumes versus COSMOS simulations, showing that both their odor experiences and resulting behaviors are quite similar.
\end{abstract}

\begin{document}

\flushbottom
\maketitle
\thispagestyle{empty}

\section*{Introduction}
Autonomous robots capable of tracking air- or water -borne chemical plumes would be beneficial for myriad applications including locating and monitoring gas leaks, wild fire outbreaks, chemical spills, nuclear contamination, and bio-hazard discharges \cite{husnain2024gas, pal2025autonomous,wang2023autonomous, macias2022optimal, soldan2012robogasinspector}. 
Developing robust plume-tracking algorithms for machines remains an open area of investigation. Ongoing efforts require overcoming the challenges of modeling the spatiotemporal odor experience for agents moving across large-scale outdoor environments, where turbulent fluctuations and the inherently intermittent nature of odor dispersal create complex patterns across space and time \cite{murlis1992odor}. Existing methods fall short either on computational expense, or in their ability to capture the natural statistics of dynamic plumes. 

One group of methods for modeling odor-plume dispersion is large‑eddy computational‑ fluid‑dynamics (CFD) solvers. These methods resolve the physics of turbulence and transport \cite{anderson1995computational, ferziger2019computational, rigolli2022learning, lionetti2023numerical}, but resolving meter‑scale eddies over large domains demands prohibitive CPU hours—even on modern hardware—making them unsuitable for rapid prototyping or real‑time control \cite{clements2024comparing}. Faster analytical alternatives such as steady Gaussian plume formulas provide smooth, time‑averaged fields that miss transient whiffs, whereas puff-based models \cite{pompy, farrell2002filament} introduce intermittency. However, their statistics diverge from field data, especially at larger scales \cite{nag2024odour}. 
Even carefully tuned CFD can misrepresent outdoor plumes when boundary conditions and turbulence parameters are imperfectly known. The new deep‑learning based time-series forecasting models \cite{rangapuram2018deep, le2020probabilistic, tang2021probabilistic, kollovieh2023predict} learn correlations from data but lack the governing physics, require massive high‑frequency datasets to capture rare whiffs, and generalize poorly outside their training regime. A more efficient strategy is to embed known physical and statistical properties directly into the model.

COSMOS addresses these gaps with a lightweight, two‑dimensional data-driven probabilistic odor simulator that models an agent's experiences when traversing through a chemical plume. COSMOS merges empirical spatial and temporal statistics with a second‑order autoregressive process to provide stochastic time series of odor encounters for a moving agent that are statistically indistinguishable from real data. Because COSMOS is a probabilistic simulator and does not model the continuous evolution of an entire plume, it operates at a fraction of the computational cost of other methods. 

Adjustable wind and source parameters allow users to tailor scenarios while preserving observed plume encounter statistics. 
COSMOS therefore enables rapid evaluation of odor‑tracking algorithms in the context of naturalistic plumes across large spatial domains to support applications including understanding animal plume tracking strategies \cite{pang2018history, ramdya2015mechanosensory}, testing algorithms before deployment on UAVs \cite{anderson2019smellicopter, shigaki2022palm}, and training reinforcement‑learning policies \cite{singh2023emergent}. 
The sections that follow detail the algorithm, validate it against outdoor plume encounter measurements and CFD, and demonstrate its utility for developing odor‑tracking strategies, concluding with broader implications and future directions.

\begin{figure}[!htb] 
\centering
\includegraphics[width=1\textwidth]{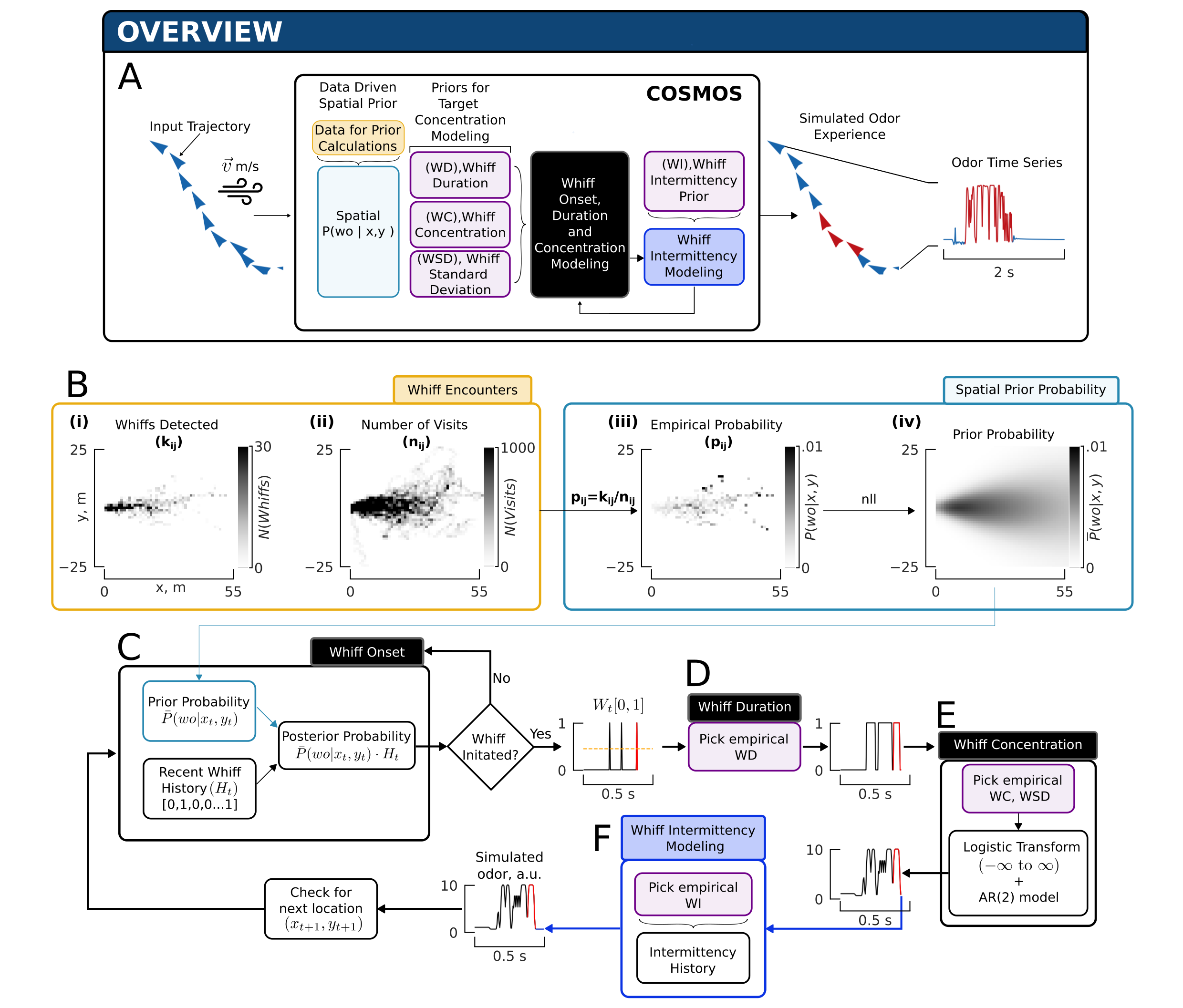}
\caption{Overview of the COSMOS pipeline for generating realistic, stochastic odor experiences. (A) High-level schematic illustrating the primary inputs (trajectory data and wind measurements) and the core algorithmic modules in COSMOS. (B) Empirical measurements of odor encounters are used to derive a spatial prior (i-iv), estimating the probability of whiff onset at each location. (C) Using recent history and spatial prior probability, the posterior probability of a whiff onset is calculated, with red representing the current whiff onset being evaluated. (D) Empirical Whiff Duration (WD) data guides transitions into or maintenance of whiff states at each time step, with the current whiff length represented in red. (E) A second-order autoregressive model combined with logistic transforms refines concentration values using empirical Whiff Concentration (WC) and Whiff Standard Deviation (WSD), ensuring smooth, realistic odor fluctuations in the simulated time series. (F) Intermittency between whiffs is modeled using empirical binned data and memory (shown in blue). At the end of an intermittency period, the next position is evaluated for the probability of a new whiff onset.}
\label{fig:algorithm}
\end{figure}

\section*{Methods}\label{sec:methods}
COSMOS takes as input a 2-dimensional agent trajectory and a time series of wind velocity vectors to simulate an odor time series for the given trajectory. To simulate the odor time series, COSMOS relies on a sequence of modules that draw from spatiotemporal odor encounter statistics from a ``template dataset"; an overview is given in  Fig. \ref{fig:algorithm} A. We first describe and demonstrate COSMOS with previously published data from an odor sensor that was carried through a desert environment with a tracer gas \cite{nag2024odour}. This template dataset contained measurements of odor concentration (ranging from 0–10 a.u.), wind velocity, and GPS locations of odor encounters, along with corresponding timestamps sampled at a fixed frequency of 200 Hz. Later, we show that our approach also works for a very different template dataset derived from computational fluid dynamics simulations. For the desert template dataset, odor concentrations exceeding a threshold of 4.5 a.u. were classified as ``whiffs," whereas measurements below this threshold were termed ``blanks." To analyze these odor encounters, we transformed the coordinate frame to align with the plume's centerline, called the streakline. In our transformed coordinate space, the \(x\)-axis represents the distance along the streakline, and the \(y\)-axis captures the perpendicular distance from the streakline. The ideal streakline was constructed by integrating wind velocity vectors over time, providing a physically grounded reference frame for mapping odor encounters. This spatial framework forms the basis for our spatiotemporal probability model, which we extend in subsequent sections to incorporate temporal dynamics.

\subsection{A Spatial Probability Map Derived from Real-World Data}
From the time series of odor measurements, a binary flag was computed to indicate the onset of ``whiffs,'' marking the start of a whiff event at the corresponding $(x, y)$ location. To transform these discrete observations into a spatial probability field, the spatial domain was discretized into a uniform grid along the $x$ and $y$ directions. The spatial domain was discretized into a uniform 50 × 50 grid, resulting in bins with dimensions of approximately 1.1 × 1.0 meters covering a total area of 55 × 50 meters. Let \(i\) and \(j\) index the rows and columns of the grid, respectively. The empirical probability of whiff onset in each bin was estimated as:
\begin{equation}
\label{eq:bin_probability}
\tilde{p}_{ij} = \frac{k_{ij}}{n_{ij}}, 
\quad 
\text{for all bins where } n_{ij} > 0,
\end{equation}
where \(k_{ij}\) is the number of whiff onset occurrences in bin \((i, j)\) (Fig. \ref{fig:algorithm}B(i)), and \(n_{ij}\) is the total number of observations in that bin (Fig. \ref{fig:algorithm}B(ii)). Bins with zero observations were assigned a default probability of zero. These empirical probabilities \(\tilde{p}_{ij}\) provide a discrete representation of the likelihood of whiff onsets across space (Fig. \ref{fig:algorithm}B(iii)). To obtain a smooth, continuous representation of the whiff onset distribution, we modeled the spatial probability of whiff onsets as a modified Gaussian plume (refer Sec. \ref{sec:gaussianplumemodel} for mathematical formulation), inspired by Farrell et al. \cite{farrell2002filament}. This final spatial probability field will be called \(\bar{P}(wo|x,y)\), as seen in Fig.\ref{fig:algorithm}B(iv), provides a smooth and continuous representation of the probability of whiff onset that preserves the statistical properties of the original empirical probabilities in Eq.\ref{eq:bin_probability} while ensuring physical plausibility through the Gaussian plume model structure.

\subsection*{Data Driven Temporal Dynamics}
Relying solely on the spatial probability of whiff onsets derived in the previous section did not yield temporal sequences with similar temporal statistics as observed in the template dataset. To address this limitation, we developed an approach that uses empirical values of whiff durations, concentrations, and intermittency values that are randomly selected from the template dataset based on the spatial location of whiff onsets. After whiffs and their subsequent blanks have ended, COSMOS returns to using the spatial probability map to ensure that whiffs are only initiated within the plume envelope. The concentrations resulting from this process are then filtered to ensure smooth transitions between states whiffs and blanks. This continuous cycle (Fig. \ref{fig:algorithm}C-F) allows us to generate unique odor time series, but with statistics that closely match the spatial, concentration, and temporal statistics of the template dataset.

\subsubsection*{Whiff onset is determined by a posterior spatial probability}
Relying solely on the spatial probability field to determine the onset of whiffs sometimes led to unrealistic temporal patterns of encounters. To tune the sparsity of whiff encounters, we multiply the spatial probability with a temporal memory term, resulting in a posterior probability:

\begin{equation}
    P_t(wo) = \alpha \cdot \bar{P}(wo|x_t, y_t) \cdot H_t,
\end{equation}
where \(\alpha\) is a constant that tunes the density of the whiffs generated, and $H_t$ introduces temporal dependence based on recent whiff history:

\begin{equation}
    H_t = \left(1 + N_W\right)
    \cdot
    \begin{cases}
    1.5, & \text{if } t_{W} > 50, \\
    1.0, & \text{otherwise,}
    \end{cases}
\end{equation}

where \(N_W\) is the count of whiff onset events in the last 20 time steps and \(t_W\) is the time elapsed since the most recent whiff. The term $(1 + N_W)$ increases probability when recent whiffs have occurred to represent a dense odor packet is encountered. If there is no whiff experienced for over 50 time steps, a small constant of 1.5 is multiplied so $H_t$ does not remain at 1 for prolonged period.
The initiation of a whiff at time step $t$ is determined by comparing the posterior probability $P_t(wo)$ to a random value drawn from a uniform distribution, where if $u<P_t(wo)$, initiate whiff and \(u \sim \text{Uniform}(0, 1)\).

\subsubsection*{Whiff durations are picked from empirical values}
When a whiff is initiated, as seen in red in Fig. \ref{fig:algorithm}D, $W_t$=1, the whiff duration (WD) is selected from the empirical distribution of whiff durations from the template dataset (e.g. Fig. \ref{fig:whiffstatgrid}A(i)). To account for the fact that the whiff duration is a function of location within the plume, we first determine the spatial bin that corresponds to the location of the whiff onset $(x_t,y_t)$, and then randomly choose an empirical whiff duration from all the durations that lie in that bin. If there are no duration in the bin, we use the mean of all empirical whiff durations which is 0.1s for HWS candidate dataset. The selected duration value determines exactly how many time steps the agent will remain in the whiff state before transitioning to intermittency.

\subsubsection*{Whiff concentrations are picked from empirical values, and filtered}
At the onset of a whiff, the location $(x_t,y_t)$ is used to pick the mean whiff concentration (WC) and whiff standard deviation (WSD) values from the corresponding spatial bin in the template dataset (e.g. Fig. \ref{fig:whiffstatgrid}A(ii-iii)). To ensure that odor concentrations smoothly transition between whiff and no-whiff states, while also simulating realistic sensor bounds, we filter the odor concentration via a two-step process using logistic transformations and autoregressive modeling as shown in Fig. \ref{fig:algorithm}E. The logistic transform converts concentration values from bounded sensor space to unbounded logit space. The autoregressive function applies a smoothing filter and distance-dependent noise to the concentration values, with noise magnitude decreasing with distance from the source. Finally, we invert the logistic transform to bring the concentration values back into the bounded space. When a whiff begins ($W_t=1$), the simulator identifies the corresponding empirical bin and a WC and WSD value is selected randomly. The WC value becomes the target concentration for the duration of the whiff, and the WSD scales the noise in the autoregressive process - hence higher WSD produce more variable odor concentrations throughtout the whiff. The same process is used to choose concentration values in the absence of a whiff, except that the WC and WSD values are chosen from no-whiff states from the template dataset.(refer to Sec.\ref{sec:concentration_modeling} for mathematical formulation and Fig.\ref{fig:whiffstatgrid}B(i-iv) for stepwise outputs)

\subsubsection*{Whiff intermittencies are picked from empirical values}
The hybrid empirical and probabilistic duration and concentration modeling described thus far effectively simulates individual whiff states. To preserve realistic temporal gaps between whiffs in our simulated odor time series, we employ an adaptive intermittency sampling strategy that considers both spatial location and recent whiff history. When a whiff ends, as seen at the end of the time series from Fig. \ref{fig:algorithm}E, the simulator identifies the spatial bin corresponding to the location $(x_t,y_t)$ and draws a random value from the empirical distribution of intermittencies corresponding to that bin (e.g. Fig. \ref{fig:whiffstatgrid}A-iv). When no template data exists for a spatial bin, we use the median of all intermittency values. This value sets the minimum number of time steps before the next whiff is allowed to start.

We found that this empirical approach sometimes led to unnaturally long bursts of whiffs, i.e. too many whiffs in a given time period. This stems from the fact that many of the observed intermittencies are quite short because turbulent plumes often contain groups of small puffs located near each other within a dense ``packet", interspersed with regions of clean air \cite{celani2014odor}. When an agent is moving through such intermittent plumes their experience will manifest as sequences that include regions of clean air, individual whiffs, and short bursts consisting of a sequence of rapid encounters. Importantly, these bursts typically do not persist for extended periods. To reproduce similar statistics with COSMOS, we retain a memory of the seven most recent intermittency values generated during the simulation. If more than half of these recent intervals were below 0.05 seconds, the algorithm finds the median of all values in the current grid cell and samples from values below this median. This helps to prevent longer than natural bursts of odor encounters. Otherwise, it samples randomly from the complete set of empirical values for that cell. 

At the end of the empirically selected whiff intermittency, COSMOS returns to the spatial posterior probability step to determine when the next whiff is initiated. This ensures that whiffs are only initiated within the spatial extend of the plume, rather than at the end of every intermittency. All these modules together constitutes COSMOS algorithm which is tunable with the hyper parameters as described in Sec. \ref{sec:cosmosParameters}.

\section*{Results}\label{sec:results} 
In this section we present and discuss the performance of COSMOS for data from three outdoor environments \cite{nag2024odordata}, as well as for a dataset derived from computational fluid dynamics simulations \cite{rigolli2022learning}. Finally, we show how COSMOS can be used to develop plume tracking algorithms.

\subsection*{COSMOS Can Simulate Real-World Spatiotemporal Odor Statistics}\label{sec:sectionhws} 
In this section, we utilized previously published data that was collected outdoors in the Black Rock Desert by carrying an odor sensor in the vicinity of a tracer gas \cite{nag2024odordata, nag2024odour}. The data consisted of three synchronized time series: odor concentration, GPS locations relative to the source, and wind vectors. We split the data into two segments, a ``high wind speed" (HWS) dataset corresponding to wind speeds ranged between 3.5 m/s and 6.5 m/s, and a ``low wind speed" (LWS), with wind speeds between 0 and 3.5m/s. First we discuss the HWS scenario. 

\begin{figure}[!hbt] 
\centering
\includegraphics[width=1\textwidth]{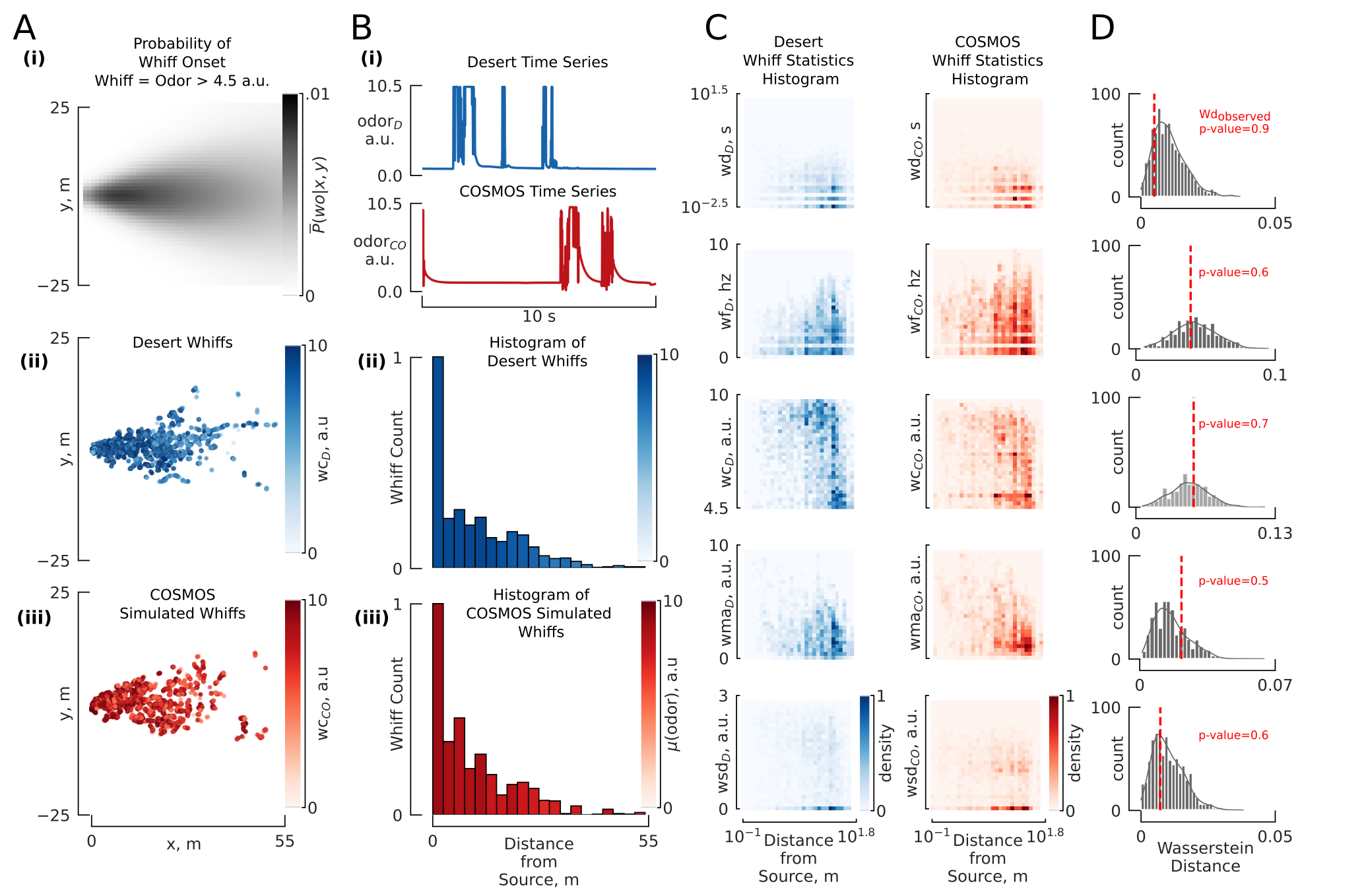}
\caption{COSMOS generates odor experiences with statistics that closely match the statistical characteristics and distributions observed in real data. Throughout the figure blue shows empirical data from the HWS dataset collected in the Black Rock Desert, whereas red corresponds to the COSMOS simulation results. Quantities with a subscript $D$ represents desert data, and with subscript $CO$ represents COSMOS. (A-i) Spatial data driven heatmap representing probability of whiff onset. (A-ii) Actual odor encounter locations from the HWS desert experiment \cite{nag2024odour}. (A-iii) COSMOS simulation of an odor experience for same trajectory as in A-ii. (B-i) Time series of odor concentration for actual and simulated odor. (B-ii) and (B-iii) presents histograms of whiff count distributions (peak normalized between 0 to 1) and average odor concentrations as a function of distance from the source for the real and simulated data, respectively. (C) Two-dimensional histograms comparing whiff statistics with distance from the source for both real and simulated data, highlighting key metrics identified in \cite{nag2024odour}. The histograms were normalized between 0 to 1. (D) Wasserstein distance distributions between real and simulated whiff statistics, bootstrapped 1000 times. The red dotted lines indicate observed Wasserstein distance values, and the p-values quantify the similarity between real and predicted distributions, higher p-values shows stronger similarity in the distributions.}
\label{fig:figure2}
\end{figure}

We used these data to generate the spatial probability map (Fig. \ref{fig:figure2}A), and whiff concentration, duration, and intermittency distributions needed by COSMOS. Fig. \ref{fig:figure2}A-ii shows the actual whiffs experienced along the trajectory, whereas Fig. \ref{fig:figure2}A-iii shows the COSMOS simulated whiff experience for the same trajectory. Although both have similar spatial distributions, differences in whiff locations and durations arise due to the stochastic nature of the simulator. This can also be seen in the time series plots in Fig. \ref{fig:figure2}B-i.

COSMOS was able to generate distributions for the number of whiffs and average concentrations as a function of distance from the source that closely matched the true data (Fig. \ref{fig:figure2}B-iii). To further evaluate the ability of COSMOS to replicate natural odor statistics, we analyzed five whiff statistics that were previously found to be correlated with odor source distance \cite{nag2024odour}. Specifically, we compared 2D histograms of whiff duration (WD, s), whiff frequency (WF, hz), whiff concentration (WC, a.u.), whiff moving average (WMA, a.u.), and whiff standard deviation (WSD, a.u.) as a function of source distance (Fig. Fig. \ref{fig:figure2}C). For each of these statistics, the COSMOS simulation yielded similar distributions compared to the real data. 

\begin{figure}[!ht] 
\centering
\includegraphics[width=1\textwidth]{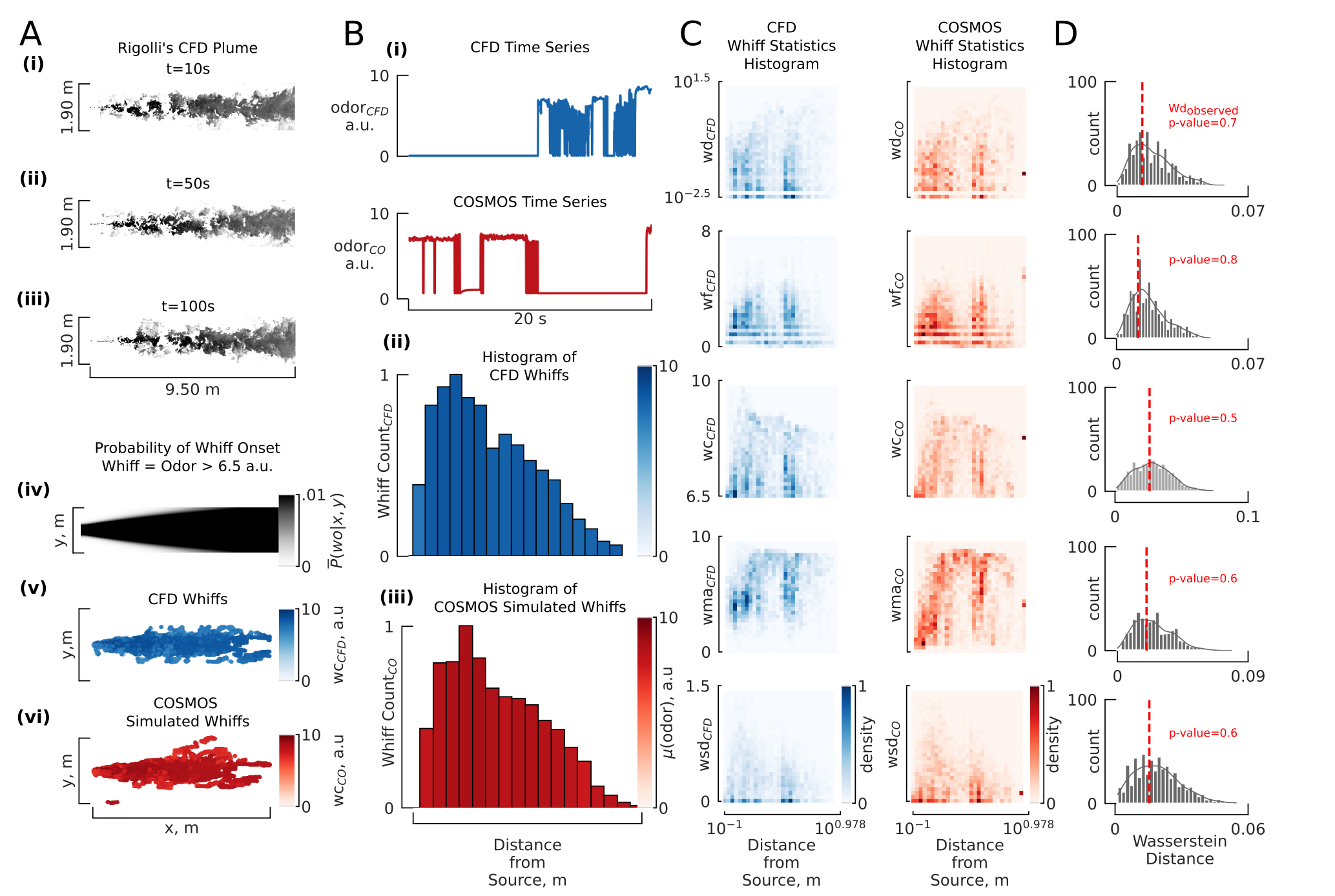}
\caption{COSMOS simulations provide odor experiences with statistics that closely match those from a dataset generated from a computational fluid dynamics (CFD) simulation \cite{rigolli2022learning}. Throughout the figure, blue represents data from the CFD dataset, while red corresponds to simulation results from COSMOS. Quantities with a subscript $CFD$ represents the CFD data, and with subscript $CO$ represents COSMOS. (A-i-iii) Snapshots of the CFD simulated plume at different time steps \cite{rigolli2022learning}. (A-iv) Data driven heatmap representing probability of whiff onset. (A-v) Actual odor experience for an agent moving through the CFD simulation. (A-vi) Simulated odor experience for same trajectory as in A-v. (B-i) Example time series traces of actual and predicted odor. (B-ii) and (B-iii) presents histograms of whiff count distributions (peak normalized between 0 to 1) and average odor concentrations as a function of distance from the source for the real and simulated data, respectively. (C) Two-dimensional histograms comparing whiff statistics (same statistics as in\ref{fig:figure2}) with distance from the source for both real and simulated data. The histograms were normalized between 0 to 1. Wasserstein distance distributions between real and simulated whiff statistics, bootstrapped 1000 times as in previous section and Fig. \ref{fig:figure2}. The red dotted lines indicate observed Wasserstein distance values.}
\label{fig:rigolli}
\end{figure}

To quantify the similarity, we calculated the Wasserstein distance \cite{kantorovich1942translocation} between the 2D histograms (Fig. \ref{fig:figure2}D, red dashed line). To assign statistical significance to the observed Wasserstein distances, we used a resampling test. We first bootstrapped null distributions. For each bootstrap iteration, we combined the real and simulated data and randomly drew samples into two groups and calculated the corresponding Wasserstein distance. Repeating this process 1,000 times generated a null distribution of bootstrapped Wasserstein distance values (Fig. \ref{fig:figure2}D, gray histograms). The position of the observed Wasserstein distance within this distribution corresponds to the probability that the observed Wasserstein distance was purely due to chance (i.e. a p-value), and provides insight into the similarity of the real-world and simulated data. A higher p-value indicates that the observed Wasserstein distance is consistent with the variability of the data, reflecting greater similarity between the distributions, whereas a lower p-value suggests dissimilarity. Our analysis revealed that all the whiff statistics were highly similar. We also conducted similar analyses on the LWS scenario from the same desert environment (Fig. \ref{fig:figurelws}), as well as a dataset collected with the same tracer gas, but in a forest environment (Fig. \ref{fig:figureforest}). Both of these datasets were also previously described in Nag et al. \cite{nag2024odordata}, and COSMOS was able to replicate similar statistics to these datasets with similar success to the HWS dataset.

\subsection*{COSMOS is Scalable and Can Learn Other Computational Simulators}\label{sec:sectionrigolli}
To evaluate the adaptability of COSMOS to other scenarios and data types, we tested it on a substantially different type of dataset: an odor plume simulated using computational fluid dynamics (CFD) \cite{rigolli2022learning, anderson1995computational} (Fig. \ref{fig:rigolli}A-(i-iii)). Because COSMOS uses data in the form of an odor experience over time for an agent trajectory, we first had to convert the plume simulation into this format. 
To ensure consistency and maintain natural movement, we used the same trajectory from the previous section, but scaled to align with the smaller spatial domain of the CFD data. Additionally, the odor concentrations in the CFD data were normalized to the same range of 0 to 10 as used in previous sections. Similar to the desert dataset, we constructed a spatial heatmap of whiff onset probabilities for the CFD data (Fig.~\ref{fig:figure2}A-iv). Because of the different dynamics between the desert and CFD data we chose a slightly different threshold for defining whiffs: 6.5 a.u. instead of 4.5 a.u., as we found this value did a good job of preserving the whiff dynamics in the CFD data. To evaluate whether COSMOS accurately replicated the odor experience we conducted the same analyses and statistical tests as with the HWS dataset, and found good agreement between the true whiff statistics and those simulated by COSMOS (Fig.\ref{fig:rigolli}B-(ii-iii), C-D).

\begin{figure}[!ht] 
\centering
\includegraphics[width=1\textwidth]{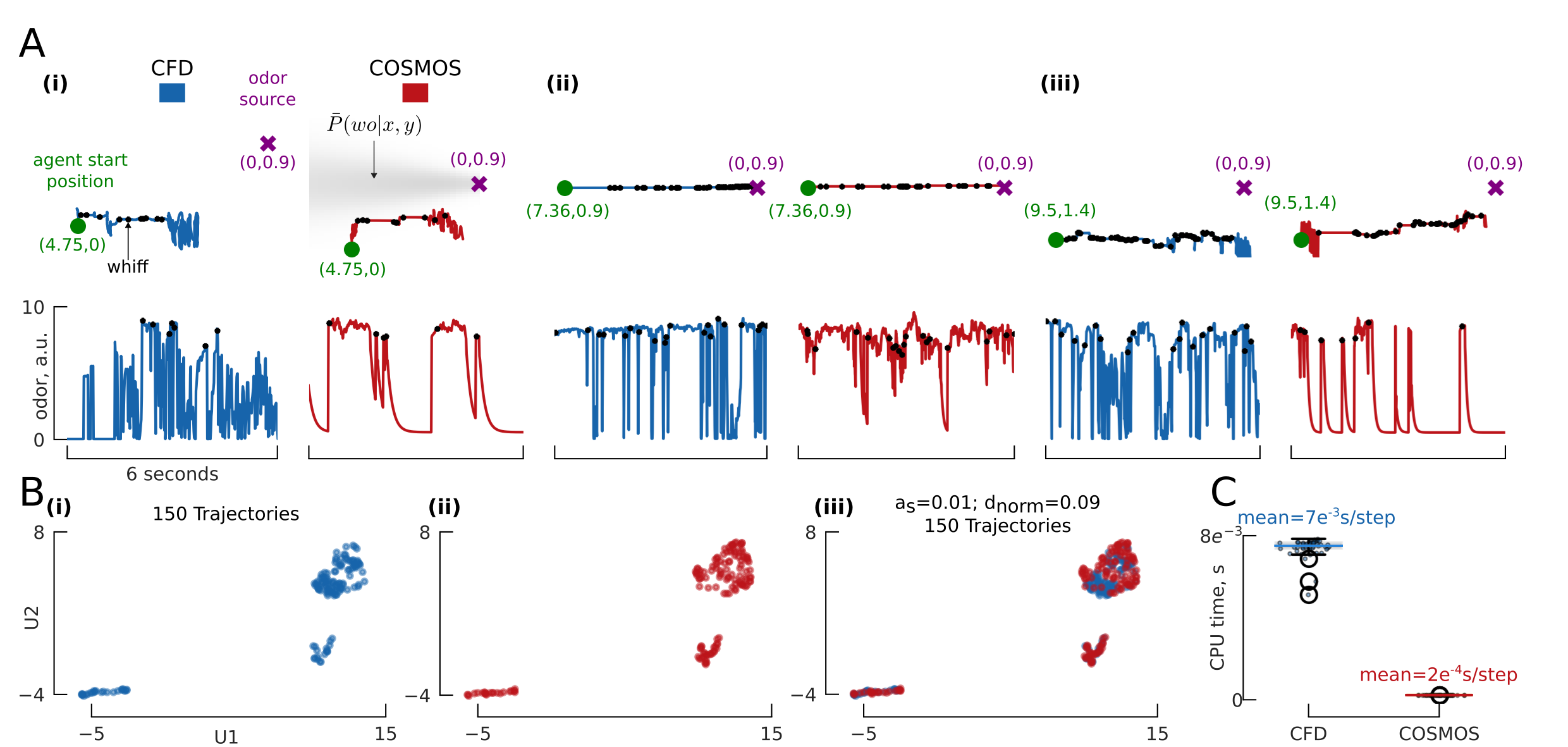}
\caption{COSMOS can be used to test odor navigation strategies. To demonstrate the application of COSMOS we used it to provide odor experiences to agents programmed with a cast and surge strategy starting from 150 different locations. Throughout the figure, blue represents simulations that used the CFD data to provide odor experiences for the moving agents, and red represents simulations that used COSMOS to provide odor experiences. (A) Representative trajectories with 6-second snippets of odor concentration experienced during these trajectories. Black dots indicate locations where whiffs were detected. In A(i), the green dot indicates the starting location of the agent, purple 'x' indicates the source position, and blue and red show the agent's trajectory. The grey scale $\bar{P}(wo|x,y)$ represents the odor probability field. (B (i-iii)) UMAP projections of multiple trajectory features, calculated from 150 simulated trajectories using either the CFD data or COSMOS simulator. The clusters show substantial overlap with a silhouette score ($a_s$) = 0.01 and a normalized centroid distance ($d_{norm}$) = 0.09 for 150 trajectories. (C) CPU time measured for 150 trajectory simulations normalized by number of steps taken in each simulations shows that COSMOS is 35 times faster than reading odor experiences from the CFD data.}
\label{fig:tracking}
\end{figure}

\subsection*{Developing Agent-Based Tracking Algorithms using COSMOS}\label{sec:sectiontracking}
One of the key applications of the COSMOS simulator is providing a framework for the development and testing of odor tracking algorithms for artificial agents. To demonstrate this application, we used a simple cast-and-surge algorithm \cite{pang2018history} and compared an agent's behavior when using either CFD based odor simulator as described in the previous section (Sec. \ref{sec:sectionrigolli}), or odor experiences derived from the COSMOS simulator built on that CFD data. 

To quantitatively compare the odor experienced by the agent in the trajectories generated by the CFD odor simulator and COSMOS, we simulated 150 odor based surge and cast agent trajectories per simulator(Fig.~\ref{fig:tracking}A). For each trajectory, we calculated several kinematic features of the trajectories (refer \ref{sec:agentTrackingParameters}). All features were normalized to ensure consistent scaling between simulators (see supplemental, Sec.~\ref{sec:trackingmaths}). The normalized features were processed using the ``tsfresh" package \cite{christ2018time} with $MinimalFCParameters$ to calculate basic statistical features (such as sum, median, mean, length, standard deviation, variance, root mean square, minimum, maximum, absolute maximum), which gave us 110 features for our kinematic features (\ref{sec:agentTrackingParameters}). To visualize this 110-dimensional space, we projected the feature space into 2-dimensions using Uniform Manifold Approximation and Projection (UMAP)\cite{mcinnes2018umap, mcinnes2018umap-software}. The resulting visualization (Fig.~\ref{fig:tracking} B(i)-(iii)) shows the distribution of trajectories where blue corresponds to CFD and red corresponds to COSMOS. The low silhouette score ($a_s = 0.01$) and small normalized centroid distance ($d_{norm} = 0.09$) indicate that agents simulated with the CFD data and COSMOS exhibited similar plume tracking behavior, as expected since the underlying odor encounter statistics were already shown to be similar.

Finally, we compared the computational cost associated with using either the CFD data or COSMOS, and found COSMOS to be much faster (Fig. \ref{fig:tracking}C). To quantify the computational cost, we simulated agents for 150 random start positions, and recorded the time required for every simulation to reach the source location and normalized the time by number of steps taken. To ensure a fair comparison across the two methods, we implemented the data extraction techniques inspired by the HDF file handling standard \cite{HDF5_CFD} example BD-CATS\cite{patwary2015bd}, specifically their approach to spatial indexing and parallel I/O when handling large scientific datasets stored in MAT files (refer to methods for computer specifications). On average, it was 35 times faster to extract odor experience from COSMOS than from CFD simulator with standard file loading techniques.

\section*{Discussion}\label{sec:discussion}
In this study, we introduced COSMOS—a Configurable Odor Simulation Model Over Scalable Spaces—that provides a computationally efficient method for simulating realistic odor experiences at large spatial scales. Earlier methods such as CFD simulations \cite{rigolli2022learning} and Lagrangian puff-based approaches like Pompy \cite{pompy, farrell2002filament}, have been limited to small spatial scales and can fail to capture real-world statistical trends. Our results demonstrate that COSMOS effectively captures the complex statistical properties of odor encounters while dramatically reducing computational overhead.  The strong concordance between empirical observations from desert field tests and COSMOS simulations—validated through Wasserstein distance metrics—confirms that our approach faithfully reproduces critical statistical features of odors encountered in natural plumes, including whiff frequency, duration, and concentration profiles. 

The computational efficiency of COSMOS, which processes data approximately 35 times faster than reading and analyzing data matrices from CFD plume simulators, enables rapid prototyping and large-scale evaluation of odor-tracking algorithms. This efficiency does not come at the expense of behavioral realism, as evidenced by the similar trajectory patterns exhibited by agents navigating both CFD and COSMOS-simulated environments. The low silhouette scores and normalized centroid distances between trajectory feature clusters confirm that agent behavior remains consistent across simulation methods, suggesting that COSMOS preserves the essential characteristics that drive navigation decisions.

The stochastic nature of COSMOS makes it particularly well-suited for reinforcement learning applications. Singh et al.  \cite{singh2023emergent} demonstrated reinforcement learning for plume tracking using a puff based odor simulator which could be replaced with COSMOS, potentially reducing computational costs while preventing policy overfitting through stochastic variability in the generated experiences.

Despite these advances, limitations remain. COSMOS currently operates in two dimensions, which does not capture the vertical dynamics of odor dispersal in three-dimensional environments. Extending COSMOS to 3D would primarily require updating the spatial onset distribution to a 3D Gaussian plume model and implementing either fully 3D or radially symmetric bins. While technically feasible using CFD datasets like Rigolli's, our focus on outdoor deployments is currently constrained by the availability of suitable 3D template datasets. The underlying statistical approach would remain fundamentally the same. COSMOS does not account for the relative movement speed between an agent and the plume. Results will be less realistic for cases where the simulated trajectory has substantially different movement speeds compared to the template dataset. This is because the statistics of odor encounters depend not only on spatial location but also on how quickly an agent traverses through odor packets. Slow-moving agents may experience prolonged exposure to individual odor patches, whereas fast-moving agents might encounter rapid succession of distinct odor packets or may not experience any odor at all. Future implementations could incorporate scaling factors that adjust temporal statistics based on relative movement speed.

The societal benefits of this work extend to applications such as environmental monitoring, hazardous material detection, and search-and-rescue operations using UAVs, where more efficient odor-guided robots could significantly improve public safety and emergency response capabilities with minimal foreseeable negative impacts due to the narrow application scope.

\section*{Acknowledgments}\label{sec:acknowledgements}
This work was partially supported by funding from AFRL (FA8651-20-1-0002), AFOSR (FA9550-21-0122), the NSF AI Institute in Dynamic Systems (2112085), and NSF EFRI-BRAID-2318081.

\section*{Author contributions}
A.N. and F.v.B. designed research; A.N. and F.v.B.  performed research; A.N. and F.v.B. contributed new reagents/analytic tools; A.N. analyzed data; and A.N. and F.v.B.  wrote the paper.

\bibliography{main}

\newpage
\section*{Supplementary Material}
\subsection{Gaussian Plume Model}\label{sec:gaussianplumemodel}
The spatial model was modeled like a smooth gaussian field. The model can be represented as

\begin{equation}
\label{eq:p_plume}
p(x,y) = 
\begin{cases}
A 
\exp\!\Bigl(-\frac{(y-y_0)^2}{2\,\sigma_y(x)^2}\Bigr)
\,\exp\!\Bigl(-\lambda\,R(x)\Bigr), & \text{if } x \geq 0 \\
0, & \text{if } x < 0
\end{cases}
\end{equation}

where $A$ is the amplitude controlling the maximum probability and $R(x)$ is the ramp function defined as: 
\begin{equation}\label{eq:ramp}
R(x) = 
\begin{cases} 
x, & \text{if } x > 0, \\
0, & \text{if } x \leq 0.
\end{cases}
\end{equation}

The first exponential term models the cross spread perpendicular to the streakline, with a distance-dependent width $\sigma_y(x)$. The second term captures the decay of probability with distance from the source along the streakline, controlled by the parameter \(\lambda\). This formulation captures the dynamic dispersion observed in natural odor plumes \cite{moore1991spatial}, with characteristic decay along the plume centerline and spreading in the cross-wind direction.  The width of the Gaussian distribution grows with distance and is given by:
\begin{equation}
\label{eq:sigma_y}
\sigma_y(x) = \sigma_{y,0} + d_y R(x)^{0.8},
\end{equation}
where $\sigma_{y,0}$ are baseline spreads, and $d_y$ is a distance-scaler controlling how the spread grows with increasing $x$. The power-law scaling reflects the sub-linear growth of plume width typically observed in turbulent flows, where theoretical predictions range from square root to linear dependence depending on atmospheric conditions and source characteristics (\cite{fischer2013mixing}, chapter 9). Our chosen exponent of 0.8 aligns with empirical observations of turbulent plume dispersion as described by \cite{vergassola2007infotaxis}.

We used a maximum likelihood approach to identify model parameters that best match our discrete empirical probabilities. We treated the discrete observations as empirical means of Bernoulli trials, where each bin \((i,j)\) has \(k_{ij}\) whiffs out of \(n_{ij}\) total observations. For a given set of parameters:
\begin{equation}
\label{eq:param_set}
\theta = \{A, x_0, y_0, \sigma_{y,0}, d_y, \lambda\},
\end{equation}
the predicted probability for a bin centered at coordinates \((x_i,\,y_j)\) is \(p(x_i,\,y_j)\). The likelihood of observing \(k_{ij}\) whiffs out of \(n_{ij}\) total observations in bin \((i,j)\) is given by the binomial probability:
\begin{equation}
\label{eq:likelihood}
L 
= 
\prod_{i,j} 
\Bigl[p(x_i,\,y_j)\Bigr]^{k_{ij}}
\Bigl[1 - p(x_i,\,y_j)\Bigr]^{\,n_{ij}-k_{ij}},
\end{equation}
and the log-likelihood is:
\begin{equation}
\label{eq:log_likelihood}
\ell 
= 
\sum_{i,j} 
\bigl[
k_{ij}\,\ln\bigl(p(x_i,\,y_j)\bigr) 
+ 
(n_{ij}-k_{ij})\,\ln\bigl(1 - p(x_i,\,y_j)\bigr)
\bigr].
\end{equation}
To find optimal parameter values we minimized the negative log-likelihood (\(-\ell\)) by bounded optimization using limited memory Broyden–Fletcher–Goldfarb–Shanno (LGBFS) algorithm  \cite{fletcher2000practical, byrd1995limited}. The parameter search bounds were within the template dataset ranges. For the template dataset shown in Fig. \ref{fig:algorithm}(B-E), we used the following parameter ranges: \(A \in [0,1]\), fixed source location \((x_0,y_0)=(0,0)\), \(\sigma_{y,0} \in [0.2,5.0]\), \(d_y \in [0.55,2.0]\), and \(\lambda \in [0.02,1.0]\). The resulting probability model p(x,y) was then spatially smoothed with a Gaussian kernel ($\sigma=1$) to produce a smooth continuous spatial probability field without strong edges given by $\bar{P}(wo|x,y)$. The output of this spatial probability field can be seen in  Fig.\ref{fig:algorithm}B(iv).

\subsection{Concentration modeling calculations using AR(2) and logistic transform for smooth transitions}\label{sec:concentration_modeling}
\begin{figure}[!ht] 
\centering
\includegraphics[width=1\textwidth]{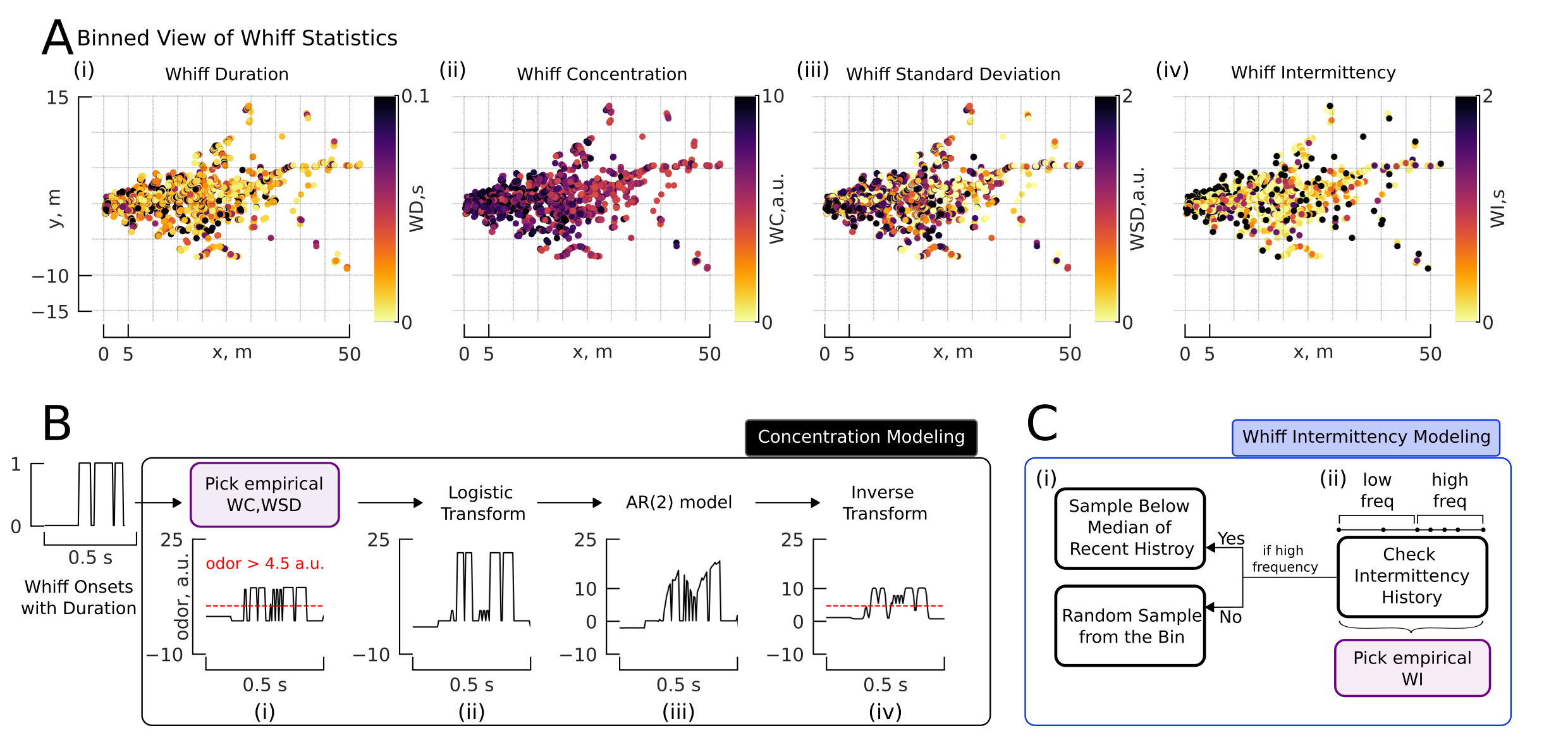}
\caption{COSMOS relies on spatially binned empirical data to determine whiff onset probabilities, whiff concentrations, durations, and intermittencies. (A) Empirical whiff characteristics experienced in Higher Wind Speed (HWS) candidate dataset binned in 5x5 meters bins (i) Whiff Concentration (WC, a.u.) (ii) Whiff Duration (WD, s) (iii) Whiff Standard Deviation (WSD, a.u.) (iv) Whiff Intermittency (WI, s).(B) Stepwise Whiff Concentration modeling and smoothing (C) Internal memory based decision making and sampling Whiff Intermittency to create similar experience as seen in HWS dense and sparse odor packets.}
\label{fig:whiffstatgrid}
\end{figure}

\begin{figure}[!ht] 
\centering
\includegraphics[width=1\textwidth]{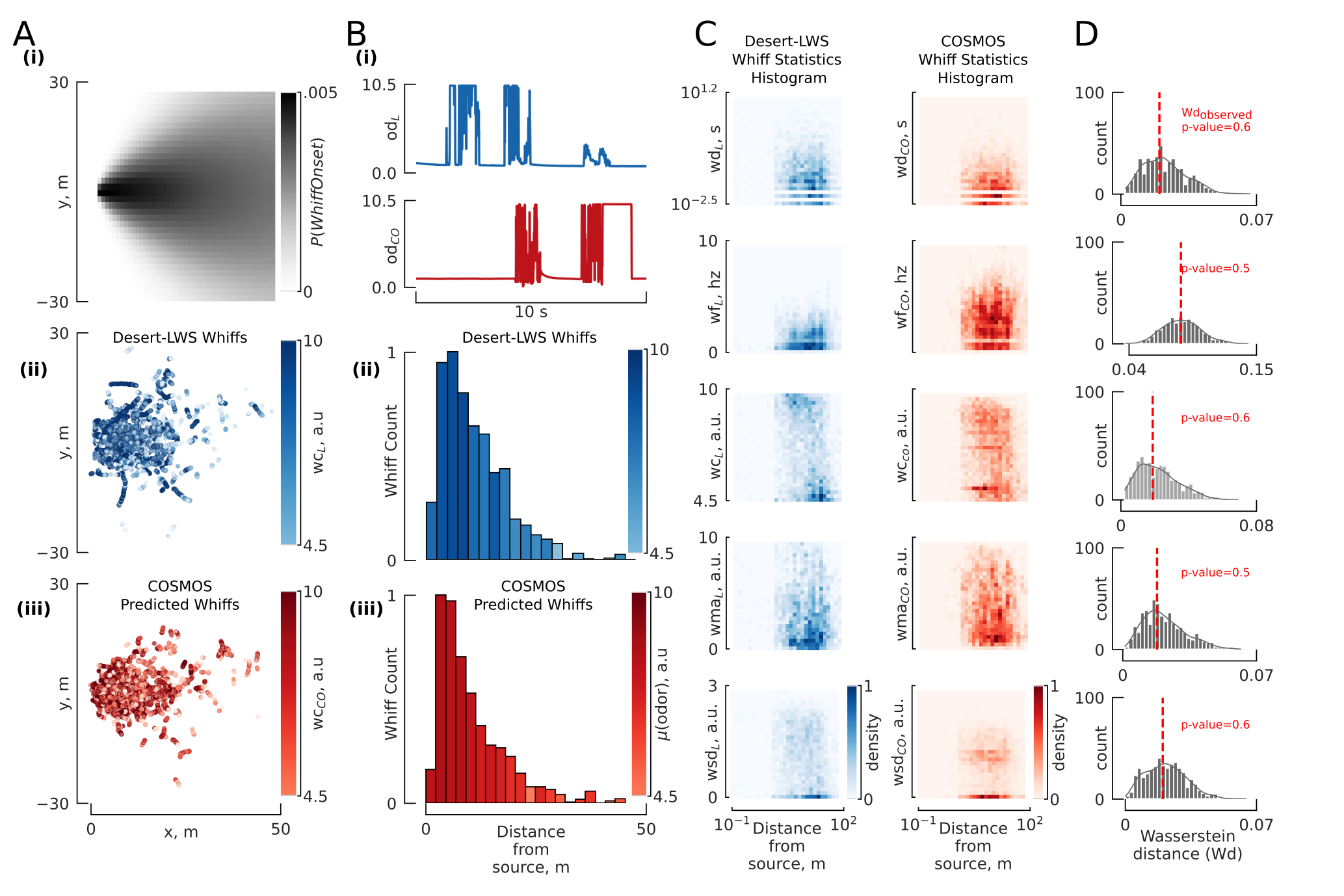}
\caption{Comparison of COSMOS simulation results to real data from the LWS dataset collected in the Black Rock Desert, plotted as in Fig. \ref{fig:figure2}. For this dataset the average wind speed was between 0 to 3.5m/s, and exhibited greater directional variability than the HWS dataset \cite{nag2024odour} (A-i) Spatial data driven heatmap representing probability of whiff onset. (A-ii) Actual odor experience in LWS dataset. (A-iii) Simulated odor for same trajectory as in LWS dataset. (B-i) 10 second sample of odor concentration time series of actual and simulated odor. (B-ii) and (B-iii) presents histograms of whiff count distributions (normalized between 0 to 1) and average odor concentrations as a function of distance from the source for the real and simulated data, respectively. (C) Two-dimensional histograms comparing whiff statistics (whiff duration (WD), whiff frequency (WF), whiff concentration (WC), whiff moving average (WMA), whiff standard deviation (WSD)) with distance from the source for both real and simulated data, highlighting key metrics identified in \cite{nag2024odour}. (D) Wasserstein distance distributions between real and simulated whiff statistics, bootstrapped 1000 times. The red dotted lines indicate observed Wasserstein distance values, and the p-values quantify the similarity between real and simulated distributions, higher p-values shows stronger similarity in the distributions.}
\label{fig:figurelws}
\end{figure}

\begin{figure}[!ht] 
\centering
\includegraphics[width=1\textwidth]{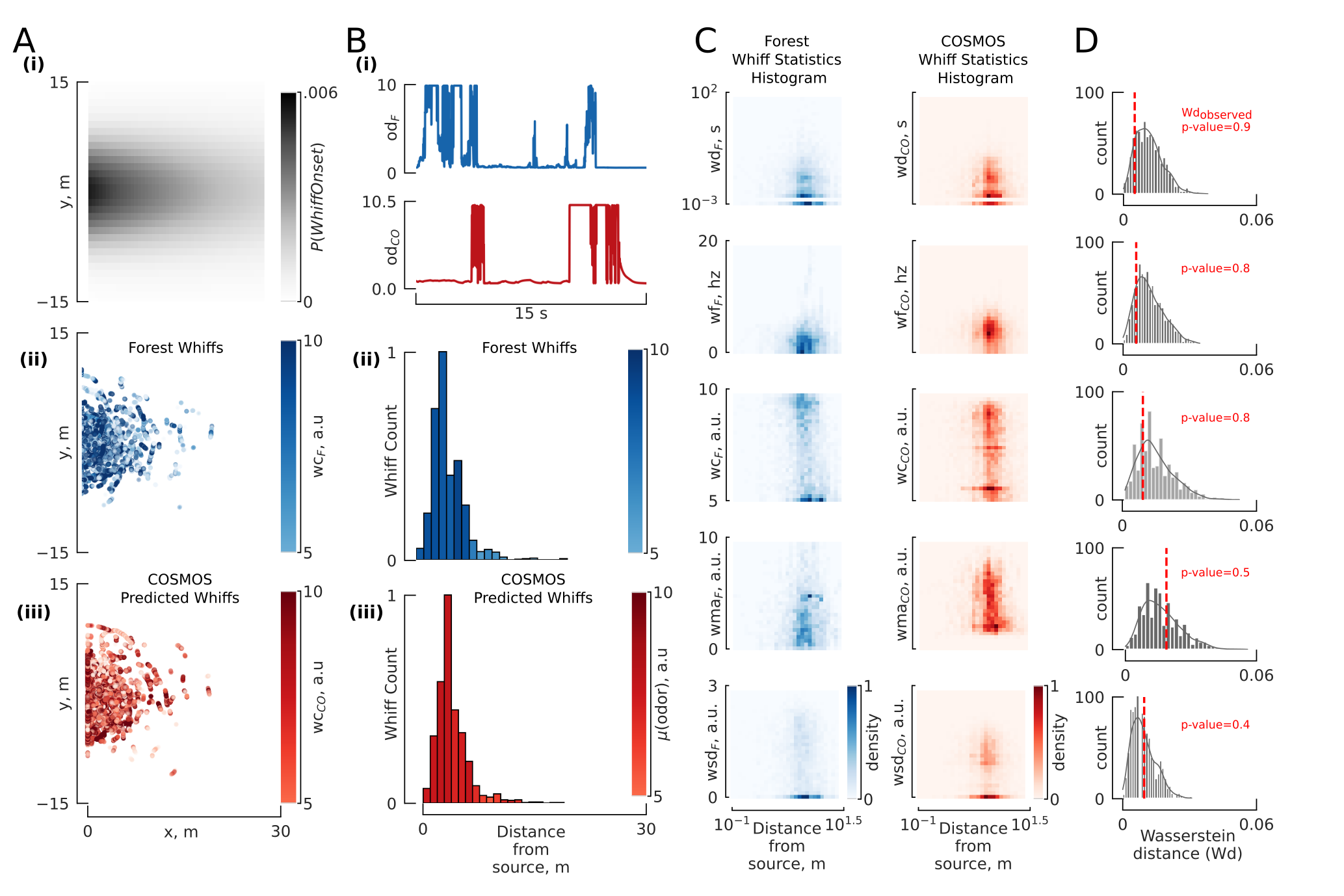}
\caption{Comparison of COSMOS simulation results to a dataset collected in a forested environment. In this dataset the average wind speed was between 0 to 10m/s with very high directional variance \cite{nag2024odour}, plotted as in Fig. \ref{fig:figure2}. (A-i) Spatial data driven heatmap representing probability of whiff onset. (A-ii) Actual odor experience in the forest dataset. (A-iii) Simulated odor for same trajectory as in forest dataset. (B-i) 15 second example of odor concentration time series of actual and simulated odor. (B-ii) and (B-iii) presents histograms of whiff count distributions (normalized between 0 to 1) and average odor concentrations as a function of distance from the source for the real and simulated data, respectively. (C) Two-dimensional histograms comparing whiff statistics with distance from the source for both real and simulated data, highlighting key metrics identified in \cite{nag2024odour}. (D) Wasserstein distance distributions between real and simulated whiff statistics, bootstrapped 1000 times. The red dotted lines indicate observed Wasserstein values, and the p-values quantify the similarity between real and predicted distributions, higher p-values shows stronger similarity in the distributions.}
\label{fig:figureforest}
\end{figure}

At each time step t, we identify a target concentration by randomly picking a concentration from the spatial bin (refer Fig. \ref{fig:whiffstatgrid}A(ii)) that contains the agent's location $(x_t, y_t)$, and call this \(C_{\text{obs}}\).For no-whiff states ($W_t$ = 0), we similarly find the nearest no-whiff measurement. For whiff states ($W_t$=1), the WSD from the corresponding bin \ref{fig:whiffstatgrid}A(iii) is used to scale the noise in the autoregressive process (Eqn. \ref{eqn:second_ar2}). Higher WSD values produce more variable odor concentrations throughout the whiff. Below are how the logistic transform and AR(2) updates are formulated. 

\paragraph{Logistic Transform:} First, we transform the bounded target concentration to an unbounded space using the logit transform,

\begin{equation}\label{eqn:logit}
z_{\text{obs}} = \ln\Bigl(\frac{C_{\text{obs}} - C_{\text{min}}}{C_{\text{max}} - C_{\text{obs}}}\Bigr),
\end{equation}
where $C_{\text{min}}$ and $C_{\text{max}}$ represents desired output bounds, for our desert template dataset these bounds follow the sensor range with $C_{\text{min}}=0$ and $C_{\text{max}}=10$. Working in this logit space prevents numerical instability in the odor dynamics and maintains natural variance in signal amplitude without getting suppressed by saturating sensor bounds. 

\paragraph{Noise Transformation:} To properly account for the non-linearity of the logit transform when calculating noise, we transform the empirical standard deviation (WSD) from concentration space to logit space:
\begin{equation}\label{eqn:noise_transform}
\begin{split}
C_{\text{upper}} = \min(C_{\text{obs}} + \text{WSD}, C_{\text{max}} - \epsilon) \\
C_{\text{lower}} = \max(C_{\text{obs}} - \text{WSD}, C_{\text{min}} + \epsilon)
\end{split}
\end{equation}
where $\epsilon$ is a small value (e.g., 0.1) to avoid numerical issues at the boundaries. We then transform these bounds to logit space:
\begin{equation}\label{eqn:logit_bounds}
\begin{split}
z_{\text{upper}} = \ln\Bigl(\frac{C_{\text{upper}} - C_{\text{min}}}{C_{\text{max}} - C_{\text{upper}}}\Bigr) \\
z_{\text{lower}} = \ln\Bigl(\frac{C_{\text{lower}} - C_{\text{min}}}{C_{\text{max}} - C_{\text{lower}}}\Bigr)
\end{split}
\end{equation}
The standard deviation in logit space is then approximated as:
\begin{equation}\label{eqn:z_std}
\sigma_{\text{noise}} = \frac{z_{\text{upper}} - z_{\text{lower}}}{2}
\end{equation}
This transformation ensures that the noise magnitude in logit space correctly corresponds to the empirical WSD in concentration space.

\paragraph{AR(2) Update:} Within this transformed space, we employ a second-order autoregressive (AR(2)) process to produce smoother changes in our simulated odor time series. At each time step, we calculate what the current concentration $z_t$ should be using:
\begin{equation}\label{eqn:second_ar2}
z_t = 
\Bigl[
\varphi_1 (z_{\text{t-1}} - z_{\text{obs}})
+
\varphi_2 (z_{\text{t-2}} - z_{\text{obs}})
\Bigr]
+
z_{\text{obs}}
+
\epsilon_t,
\end{equation}

where the coefficients are, $\varphi_1$ = 0.85, $\varphi_2$ = -0.17. The parameters were selected using Python’s \texttt{statsmodels.tsa.AutoReg} \cite{statsmodels_autoreg}, where we then fitted our logit transformed data (Eqn. \ref{eqn:logit_bounds}) to get the values of $\varphi_1$ and $\varphi_2$. In Equation \ref{eqn:second_ar2} $\varphi_1$ facilitates rapid adjustments to the current state, while $\varphi_2 < 0$ provides damping to prevent oscillations. Together, these parameters control the system's responsiveness to deviations from the target concentration. 
The term $\epsilon_t$ represents Gaussian noise drawn from $\mathcal{N}(0, \sigma_t)$, where the standard deviation $\sigma_t = \sigma_{\text{noise}} \cdot (1 + e^{-d/50})$ depends on the distance $d$ from the source and $\sigma_{\text{noise}}$ is the transformed standard deviation as described above. After each time step, we ensure that the concentration values remain within the original range by transforming the prediction back to concentration space using the inverse logit transform,

\begin{equation}\label{eqn:invlogit}
C_{\text{t}} = C_{\text{min}} + \frac{C_{\text{max}} - C_{\text{min}}}{1 + e^{-z_t}}.
\end{equation}

where $C-t$ represents the final simulated odor concentration that would be experienced by the agent at time t at location $(x_t,y_t)$.

\subsection{Agent Tracking Parameters for Tsfresh} \label{sec:agentTrackingParameters}
To enable trajectory comparison between COSMOS and CFD simulations, we calculated the following features for each agent trajectory:
\begin{itemize}
    \item Kinematics: velocity components $(v_x, v_y)$, speed $\|\mathbf{v}\|$, acceleration $\|\mathbf{a}\|$, and position relative to the source measured as crosswind distance $y_{cw}$ and upwind distance $x_{uw}$
    \item Orientation: heading angle $\theta$ and angular velocity $\omega$ (equivalent to the turn rate $\dot{\theta}$)
    \item Path characteristics: curvature $\kappa$, path length $l$, and turn rate $\dot{\theta}$ (calculated as the gradient of heading angle with respect to time)
\end{itemize}

These features were normalized across both simulation environments and processed with ``Tsfresh" to extract statistical features, which were then used for UMAP projection and trajectory clustering analysis.

\subsection{System Specification} \label{sec:computerSpecs}
All analysis were performed in a System76 Thelio Mega workstation, with 128GB memory (RAM), processor - AMD Ryzen threadripper pro 5955wx 16-cores x 32 , gpu - NVIDIA GeForce RTX 4090, running linux pop os.

\subsection{Trajectory Analysis}\label{sec:trackingmaths}
To ensure consistent scaling between simulators, we applied feature-specific normalization:
\begin{equation}
    \tilde{f_i} = \frac{f_i}{\max(|f_i|)} \quad \text{for spatial and kinematic features}
\end{equation}
\begin{equation}
    \tilde{\theta} = \frac{(\theta \bmod 2\pi)}{\pi} - 1 \quad \text{for angular measurements}
\end{equation}

The high-dimensional feature space was reduced using Uniform Manifold Approximation and Projection (UMAP), which optimizes:
\begin{equation}
    \mathcal{L} = \sum_{i,j} [p_{ij}\log(\frac{p_{ij}}{q_{ij}}) + (1-p_{ij})\log(\frac{1-p_{ij}}{1-q_{ij}})]
\end{equation}
where $p_{ij}$ represents high-dimensional similarities and $q_{ij}$ low-dimensional similarities. We use the following parameters in the UMAP projection:

\begin{itemize}
\item $n_{neighbors}=50$ for global structure preservation
\item $min\_dist=0.1$ for cluster separation
\item Euclidean metric for feature space distances
\end{itemize}

The quality of simulator comparison was quantified using two metrics:
\begin{equation}
    a_s = \frac{b(i) - a(i)}{\max(a(i), b(i))} \quad \text{(silhouette score)}
\end{equation}
where $a(i)$ is the mean intra-cluster distance and $b(i)$ is the mean nearest-cluster distance, and
\begin{equation}
    d_{norm} = \frac{\|\mu_{CFD} - \mu_{COSMOS}\|}{\frac{1}{2}(\sigma_{CFD} + \sigma_{COSMOS})} \quad \text{(normalized centroid distance)}
\end{equation}

\subsection{Code and Data Availibility}\label{sec:codeanddata}
All code is available in \href{https://github.com/arunavanag591/COSMOS}{github}. All data is available in \href{http://datadryad.org/share/6ahtoddnVD7c3Tj2zKHLjVn3GTtAj-W6zqIYu9udpL4}{data dryad}. 

\subsection{COSMOS parameters} \label{sec:cosmosParameters}
COSMOS requires the spatial odor experience model (similar to the model in Fig \ref{fig:algorithm}B(iv)) for a plume, along with whiff characteristics and an array of $(x,y)$ points for the trajectory to be tested. Beyond these inputs, COSMOS allows users to tune various hyperparameters to achieve their desired density of odor experience.
\begin{table}[!ht]
\centering
\small
\begin{tabular}{p{0.2\textwidth}p{0.15\textwidth}p{0.55\textwidth}}
\toprule
\textbf{Parameters} & \textbf{Default Value} & \textbf{Description (Tunable parameters for COSMOS)} \\
\midrule
\multicolumn{3}{l}{\textbf{Temporal Resolution}} \\
\midrule
rows\_per\_second & 200 & Controls temporal resolution of the simulation. Higher values give finer time resolution but increase computation time. \\
\midrule
\multicolumn{3}{l}{\textbf{Probability Configuration}} \\
\midrule
density\_scaler & 1 & Multiplied with the spatial odor experience model to control density or probability of whiffs experienced by the agent\\
whiff\_transition\_prob($\alpha$) & 0.85 & Probability of transitionting from a whiff or no whiff state to a whiff state \\
\midrule
\multicolumn{3}{l}{\textbf{Whiff State Parameters}} \\
\midrule
base\_odor\_level & 0.6 & Baseline odor concentration when no whiffs are present. Sets the minimum concentration floor. \\
low\_threshold & 0.05 & Threshold for considering intermittency values as "low". Affects bias in intermittency sampling. \\

\midrule
\multicolumn{3}{l}{\textbf{AR(2) Model Parameters}} \\
\midrule
ar1 ( $\varphi_1$) & 0.85 & First-order autoregressive coefficient. Controls short-term temporal correlation. \\
ar2 ( $\varphi_2$) & -0.17 & Second-order autoregressive coefficient. Controls longer-term temporal correlation. \\
\midrule
\multicolumn{3}{l}{\textbf{Memory Parameters}} \\
\midrule
lookback\_history & 50 & Number of timesteps to lookback in past to make decision. \\
history\_intermittency & 7 & Number of recent intermittency values to consider. \\
\midrule
\multicolumn{3}{l}{\textbf{Signal Processing Parameters}} \\
\midrule
window\_size & 14 & Size of rolling window for smoothing and statistics. Affects final signal smoothness. \\
sigma & 0.8 & Gaussian filter width for final smoothing. Controls high-frequency noise reduction. \\
\bottomrule
\end{tabular}
\caption{COSMOS Simulator Hyper-Parameters and Their Descriptions}
\label{tab:parameters}
\end{table}

\newpage
\subsection{COSMOS Results for LWS and Forest}

\clearpage

\beginsupplement
\renewcommand\figurename{Supplementary Figure}
\renewcommand\tablename{Supplementary Table}

\clearpage

\end{document}